# Eusprig 2006 Commercial Spreadsheet Review

Simon Murphy
Codematic Limited
50 PatriotHall
Edinburgh
UK
Simon.murphy@codematic.net

## ABSTRACT

*This management summary provides an outline of a commercial spreadsheet review process. The aim of this process is to ensure remedial or enhancement work can safely be undertaken on a spreadsheet with a commercially acceptable level of risk of introducing new errors.*

## 1    INTRODUCTION

All successful software gets maintained (Glass, 2003). A key part to maintaining and enhancing any system is understanding how things work and how the different elements interrelate. Glass also suggests that it is more difficult to understand and maintain somebody else's work than to write from scratch (Glass, 2003).

Codematic will maintain/enhance/fix anyone's spreadsheet resources, whatever their history and current state. In order to do that as a commercial venture we need to develop a reasonable understanding of a spreadsheet in a very cost effective manner. This paper outlines the process used in the bid phase of fixed price break/fix quotes. The targets are business applications, generally of a financial nature, often as part of a Management Information System for a commercial organisation. The ideas may well be applicable elsewhere.

Most work is actually Excel based but the techniques could equally be applied in other spreadsheet applications such as Open Office Calc and Gnumeric. We also recognise the significant limitations of the spreadsheet paradigm (Murphy, 2005), and will recommend migration wherever appropriate.

Zapawa (2005) suggests the real value of a spreadsheet is the value of a good decision less the value of a bad decision based on its outputs. Another measure of spreadsheet value is the cost to develop the model – this can be millions of pounds (Croll, 2005). By either measure many spreadsheets are worth looking after.

## 2    BACKGROUND

Working with spreadsheets offers some unique challenges (for example they are generally a binary format which is more difficult to manage in a source control system). Although similar to mainstream software development in some ways, spreadsheets can also be quite idiosyncratic (Murphy, 2005). One significant barrier to understanding spreadsheets in general is the very flat object hierarchy:

| Spreadsheet | Text | Relational Database | High level code |
|---|---|---|---|
| Cell | Word | Value | Instruction |
| Sheet | Sentence | Field | Line |
| Workbook | Paragraph | Record | Procedure |
| | Chapter | Table | Class/Module |
| | Document | Database | Application |

Compared to other applications spreadsheets do not have many layers, this makes a gradual drill down from overview to detail difficult. Well designed spreadsheet models will often add more levels to make comprehension easier. One very useful additional layer is the analysis block, being a group of cells that perform a single analysis step.

Another issue specific to spreadsheets is the massive breadth of uses they are put to. From shopping lists to Currency Market trading applications and pretty much everything else, spreadsheets are ubiquitous in the business world and beyond.

## 3    PAPER OVERVIEW

This paper discusses our review process, some of the tools we use and some of the reasons behind the decisions made. The primary aim of our review is to gain sufficient knowledge of a spreadsheet to:

a.  quote a fixed price to fix or enhance it
b.  Make the required changes whilst maintaining or enhancing the integrity of the model.
c.  Or recommend migration
d.  Or decline the opportunity if appropriate.

## 4    SPREADSHEET REVIEW STAGES

The following are the main stages in our review:

1.  Clarify the scope of the review – in particular the limitations.
2.  Understand the spreadsheet in the wider context of the client organisations information systems
3.  File Review
4.  Workbook level review
5.  Sheet summary
6.  Inter-sheet interactions
7.  Cell/Block level relationships

### 4.1    Scope of review

We look for insight into the abilities and attitude of the people who have been responsible for the spreadsheets development. The main question is: Is this the sort of model that can be maintained safely? Can it be maintained commercially? How important is the spreadsheet to the client? This is a superficial review, not a thorough, or complete, examination. We do not aim to 'prove' a model is correct, or fit for purpose, or even free from material error or fraud. Our aim is to be able to make the changes a client requires, at a price they are willing to pay. Proving a model is correct is a completely different activity. Our approach is based on the explicit assumption that the model we initially review is currently considered fit for purpose, or was until it broke.

The review assumes limited knowledge of the clients business or the purpose of the spreadsheet. These would be critical for a deep understanding, but not essential for most maintenance activities.

The client may provide additional documentation – this is likely to be out of date and irrelevant or even misleading.

Target times are really dependent on the complexity of the model, but half a day should be enough to quote. Significant further work may be required before the model can be safely edited.

## 4.2 Spreadsheet context

What is the spreadsheet used for? Do its outputs feed any other models? Do its outputs drive significant business decisions? Where do its inputs come from? How stable are those sources? How reliable? Does it use other technologies like databases or add-ins?

If the spreadsheet is tied closely to a back end system like a SQL Server database or Essbase, then an off-site review may be inappropriate.

## 4.3 File Review

All files from external sources are always scanned for viruses with the latest definitions before being allowed onto the Codematic network.

The first step is actually outside Excel – it is reviewing the workbook properties in particular looking for any useful custom ones, and the protection status. If the workbook requires a password to open and it is not provided the review goes no further.

The review process is fairly destructive, so is always carried out on a copy of the source file. This copy also has any workbook structure passwords, worksheet protection and VBA protection removed to ease analysis.

## 4.4 Workbook Level review

The first step in the review is safety driven again – open the workbook with Macros and VBA disabled. Review any code for security threats and for general quality. Great insight can be gained from the code, and it often helps categorise the skill and experience level of the developer. Actual ability is less important that any apparent over confidence.

The next stage is to run XLAnalyst (free version available) to get a brief summary report. The key thing about XLAnalyst is its fast (generally just a few seconds), and gives a reasonable overview of some potential issues. The commercial version also produces a list of unique formulas which is very useful for classifying the model as complex or simple.

| | | | Results | Info | Example address | Example formula | Weighting (0=N, 9= critical) |
|---|---|---|---|---|---|---|---|
| 5 | | **Workbook name:** PLDemo.xls | | | E:\data\xlec\PLDemo.xls | | |
| 6 | | **Overall Risk Rating =** 34% | | | Higher value means more chance of defects. Low risk rating is not the same as defect free. | | |
| 8 | | **Summary Potential Risk Report** | Results | Info | Example address | Example formula | |
| 9 | | **Factors suggesting a high risk of an error** | | | | | |
| 10 | ? | Circular References | Not Found | | | | 10 |
| 11 | ? | Cells Displaying A Number But Storing Text | Not Found | | | | 10 |
| 12 | ? | Mixed Formulas And Values | Found | | PhasingTable!$C$8 | =IF(MAX(C8:AP19)>2,"Warning some growths are more than do | 10 |
| 13 | ? | Formulas Evaluating To An Error | Not Found | | | | 10 |
| 14 | ? | Vlookups Expecting An Ordered List | Not Found | | | | 8 |
| 15 | ? | Hlookups Expecting An Ordered List | Not Found | | | | 8 |
| 16 | | | | | | | |
| 17 | | **Factors suggesting a significant risk of an error** | | | | | |
| 18 | ? | Links To External Workbooks | Not Found | | | | 5 |
| 19 | ? | Presence Of Very Hidden Sheets | Not Found | | | | 5 |
| 20 | ? | Hidden Rows Or Columns | Found | | PhasingTable!Col | is hidden | 3 |
| 21 | ? | "+" Construct | Not Found | | | | 3 |
| 22 | ? | Conditional Formatting | Found | | North!$E$4 | =D4+1 | 3 |
| 23 | ? | Use Of Pivot Tables | Not Found | | | | 3 |
| 24 | | | | | | | |
| 25 | | **Factors suggesting complex logical modelling** | | | | | |
| 26 | ? | Array Formulas | Not Found | | | | 8 |
| 27 | ? | Nested If Statements | Not Found | | | | 6 |
| 28 | ? | Use Of Sumif | Not Found | | | | 5 |
| 29 | ? | Use Of Database Functions (Dsum Etc) | Not Found | | | | 5 |
| 30 | ? | Use Of Indirect | Not Found | | | | 5 |
| 31 | | | | | | | |
| 32 | | | | | | | |
| 33 | ? | **Measures** Longest Formula | Within Limit | 66 | PhasingTable!$C$8 | =IF(MAX(C8:AP19)>2,"Warning some growths are more than do | 7 |
| 34 | ? | Most Complex Formula | Above Limit | 8 | Total!$P$13 | =IF(ROUND(SUM(D12:O12),0)<>ROUND(P12,0),"Cross Cast Er | 7 |
| 35 | ? | Total Number Of Formulas | Above Limit | 2,176 | | | 5 |
| 36 | ? | Total Number Of Unique Formulas | Above Limit | 13 | | | 5 |
| 37 | ? | Workbook Size | Above Limit | 132 Kb | | | 5 |
| 38 | ? | No Of Worksheets | Above Limit | 8 | | | 5 |
| 39 | ? | Total All Lines Of VBA Code | Above Limit | 290l9 | | 290 Lines In 9 Components | 8 |
| 40 | ? | Largest Formula Result | Above Limit | 2.5E+04 | Total!$P$12 | =SUM(P8:P11) | 0 |
| 41 | | | | | | | |
| 42 | | **System messages** | Found | 1 | | The following | |

| | | | | |
|---|---|---|---|---|
| 1 | **XLAnalyst Unique Formula List report - Test Date: 25 February 2006 - 20:38:41** | | | |
| 2 | Worksheet Name | Cell Address of first found | Formula <f> or Array Formula <af> (double click to Go | |
| 3 | PhasingTable | $R$5 | <f>=Q5+1 | |
| 4 | PhasingTable | $C$6 | <f>=IF(MAX(C8:AP19)>2,"Warning some growths are | |
| 5 | PhasingTable | $C$23 | <f>=MAX(C8:AP19) | |
| 6 | PhasingTable | $C$24 | <f>=MIN(C8:AP19) | |
| 7 | North | $P$8 | <f>=SUM(D8:O8) | |
| 8 | North | $Q$8 | <f>=D8*PhasingTable!$C$8*PhasingTable!Q8 | |
| 9 | North | $D$12 | <f>=SUM(D8:D11) | |
| 10 | North | $D$13 | <f>=IF(SUM(D12:O12)<>P12,"Cross Cast Error","") | |

If a model receives a high XLAnalyst score then it is likely to be harder to work with, although a good overall design can more than outweigh a high score.

If XLAnalyst throws up any linked workbooks then if possible those are traced back through the (Clients) file system using an automated tool (Internal product). Tracing the links can uncover hidden circularity where through a series of files a cell references itself. This can be a very useful context setting exercise too – One client spreadsheet linked to 34 others, 14 of which were no longer available on the network. Update inconsistencies can also be highlighted here too, where a source file has a later saved date than a file that feeds from it.

Review workbook metrics to get a sense of the type of workbook:

There are several interesting things in the following example:
1. Over 200 worksheets might suggest that potential design improvements could be made.
2. Almost a million formulas might suggest a database could be a more suitable tool for this system.
3. A very large number of blank cells in the used range would suggest some mild used range corruption, and sure enough resetting the last cell on each sheet reduced the file size from 179Mb to 69Mb, and more importantly reduced the chance of more serious corruption later.

4. Finding the clients problem could be difficult in such as spreadsheet, this could end up as a migration job rather than a find and fix job.

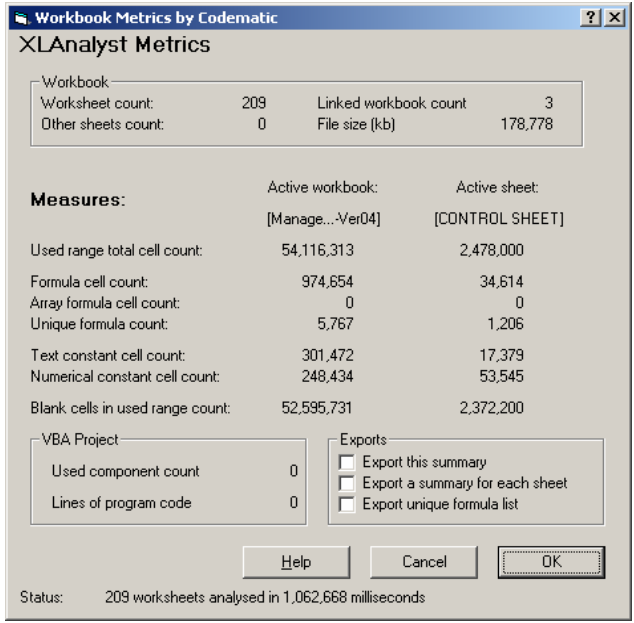

Later on the exports will be used at the sheet level to summarize and give context at that level. More review effort will be spent of those sheets with more formulas and more unique formulas.

Once the workbook level context has been assessed the analysis gradually drills into the model one level at a time.

## 4.5 Sheet Level Review

To help navigation around the model an index sheet is inserted. It contains hyperlinks to all sheets, any sheet level descriptions and the ability to group sheets together and hide and unhide them. It also makes tracking review status simple. (WorkbookStructure – available free). The workbook is reopened with macros enabled to allow the index sheet to work properly.

The metrics summaries are also reviewed to assess how the calculations are spread across the sheets.

## 4.6    Inter-sheet Interactions

The next drill level after workbook is sheets.  It is very useful to see how data flows through the worksheets in a workbook.  A clear one directional flow is a good sign, circularity or confusion often indicates a design weakness, and a more difficult review. Effective use of worksheets as modules with meaningful names makes this step very useful.  A lack of such structure renders it fairly useless.

Data flow for FLDemo.xls   (E:\data\xlac\)
Formula links only (misses spiked sums)
Named ranges and graph links not included
Diagram generated on: 25/02/2006 20:36:50
XLAnalyst-XLFlow version: xlflow47.xla

## 4.7    Cell/Block Level review

Unhide all rows and columns on all sheets, colour them as they are more likely to contain accidental overwrite errors.

Run colour mapper (available free) to map out the text, numeric values and formulas. Other tools provide more colours, but just the three seems easier to manage.  Zooming out to 25% or so, gives an overview and also causes Excel to display any range names that have been defined.

This step looks for layout logic, but also for values in the middle of formulas or text amongst numbers. If text and formulas are in clear blocks that may indicate a sound design, if they are all jumbled up the model may be harder to work with.

The next stage is to drill into the worksheets of a fresh version of the workbook and gain some understanding of what is happening at the formula level.

Map out all the cells flows in the sheet to see how the cells inter-relate. In many finance models the flow is to the right and downwards, other schemes may be appropriate. The main concern would be if the flows seemed almost random, such a model may be very difficult to modify safely.

The final part of the review is to step though some of the dependency trees. It is useful if the tool evaluates lookups and offsets and so on to show the actual cell being referenced. The aim here is to confirm the model works as expected based on the previous higher level work. Sometimes the dependency logic can be unclear, especially if some of the terminology is unfamiliar, this step can be very useful for generating questions for the client to help gain a deeper understanding of what the model does.

| | | 2005 Actuals/Forecast | | | | | | | | | | |
|---|---|---|---|---|---|---|---|---|---|---|---|---|
| | | | 4 | 5 | 6 | 7 | 8 | 9 | 10 | 11 | 12 | Total | 1 |
| £000's | | | Act | Act | Act | Act | Act | F/c | F/c | F/c | F/c | | F/c |
| **Revenue** | | | | | | | | | | | | | |
| Advertising Revenue | | | 50 | 215 | 169 | 287 | 183 | 74 | 191 | 259 | 211 | 2,497 | 269 |
| Circulation Revenue | | | 231 | 157 | 325 | 76 | 318 | 216 | 305 | 148 | 56 | 2,350 | 130 |
| Other Revenue | | | 85 | 90 | 85 | 149 | 79 | 50 | 167 | 111 | 230 | 1,440 | 87 |
| **Total Revenue** | | | 366 | 462 | 580 | 511 | 580 | 341 | 663 | 518 | 496 | 6,286 | 486 |
| **Costs:** | | | | | | | | | | | | | |
| Employment | | | 2 | 182 | 253 | 219 | 338 | 178 | 104 | 71 | 80 | 2,271 | 264 |
| Newsprint | | | 41 | 79 | 97 | 71 | 89 | 22 | 23 | 68 | 68 | 763 | 46 |
| Other printing | | | 38 | 69 | 4 | 61 | 59 | 35 | 55 | 56 | 20 | 605 | 70 |
| Depreciation | | | 38 | 20 | 40 | 8 | 102 | 7 | 78 | 38 | 110 | 565 | 29 |
| Other | | | 22 | 8 | 5 | 15 | 6 | 21 | 15 | 28 | 15 | 176 | 16 |
| **Total Costs** | | | 409 | 357 | 399 | 374 | 595 | 263 | 274 | 261 | 293 | 4,381 | 433 |
| **Net margin contribution** | | | (43) | 105 | 180 | 138 | (16) | 78 | 389 | 257 | 203 | 1,905 | 54 |
| ROS % | | | (11.6%) | 22.7% | 31.1% | 26.9% | (2.7%) | 22.9% | 58.6% | 49.6% | 40.9% | 30.3% | 11.0% |

If the spreadsheet is reasonably well designed it should be possible to summarise the purpose of each sheet in a couple of sentences. This should be added, if not already there, somewhere (easy to find and maintain) in the sheet (A1 for example) and referenced by the index sheet. This is a useful way of measuring review progress. Once all the sheets have been reviewed it should be possible to quote or decline.

Fully detailed, comprehensive cell by cell review is not part of this process, but some elements of the model may require this additional analysis if the project is undertaken.

## 5    SUMMARY

This paper has briefly summarized one commercial approach to reviewing spreadsheets to gain some level of comfort with the analysis it performs.

The approach focuses on drilling down stage by stage, in a methodological manner, into the details of the model whilst maintaining some context.

The aim is to gain sufficient understanding to work on the model and be reasonably comfortable that new errors will not be introduced.

## 6    CONCLUSIONS

It is possible using various free tools and tools included with Excel to review a spreadsheet to a level of comfort that would permit a fixed price quote to find and fix a problem or apply an enhancement.

Whilst falling far short of 'proving' the correctness, or otherwise, of a model, the approach as outlined enables remedial work to be carried out within commercially acceptable risk boundaries.